\documentclass[]{revtex4}
\usepackage{amsmath}
\usepackage{graphicx}
\usepackage{amsfonts}
\usepackage{amssymb}

\begin{document}

\title{Controlled vaporization of the superconducting condensate in cuprate
superconductors sheds light on the pairing boson.}
\author{ P. Kusar,$^{1}$ V. V. Kabanov,$^{1}$ S. Sugai,$%
^{2}$ J. Demsar,$^{1,4}$ T. Mertelj$^{1,3}$and D. Mihailovic,$^{1}$ }
\affiliation{$^{1}$Complex Matter Dept., Jozef Stefan Institute, Jamova 39, Ljubljana,
SI-1000, Ljubljana, Slovenia
$^{2}$Department of Physics, Faculty of
Science, Nagoya University, Chikusa-ku, Nagoya 464-8602, Japan
$^{3}$
Department of physics, Faculty of mathematics and physics, Jadranska 19,
Ljubljana, Slovenia
$^{4}$University of Konstanz, Physics Department,
D-78457 Konstanz, Germany }
\date{Submitted for publication : 9.Nov.2007}

\begin{abstract}
We use ultrashort intense laser pulses to study superconducting state vaporization
dynamics in La$_{2-x}$Sr$_{x}$CuO$_{4}$ ($x=0.1$ and 0.15) on the
femtosecond timescale. We find that the energy density required to vaporize the
superconducting state is $2\pm 0.8$ K/Cu and $2.6\pm 1$
K/Cu for $x=0.1$ and 0.15 respectively. This is significantly
greater than the condensation energy density, indicating that the quasiparticles share a large amount of energy with the boson glue bath on this timescale. Considering in detail
both spin and lattice energy relaxation pathways which take place on the relevant timescale of $\sim10^{-12}$ s, we rule out purely
spin-mediated pair-breaking in favor of phonon-mediated mechanisms, effectively ruling out spin-mediated pairing in cuprates as a consequence.
\end{abstract}

\maketitle

The study of nonequilibrium phenomena in superconductors has been an
important topic of condensed matter physics since the 1960's and the fact
that intense laser pulses can non-thermally destroy the superconducting
state has been known for a long time \cite{Testardi}. After the discovery of
high-temperature superconductivity in cuprates and the simultaneous rapid
development of ultrashort pulsed lasers, real-time studies of quasiparticle
(QP) dynamics became possible using pump-probe experiments \cite%
{Han,Chwalek,Albrecht,Stevens}.
 Recent developments in phenomenological modeling and new systematic experimental studies of the non-equilibrium
optical response \cite{RT,MgB2,nasiRT,NasiPRB1999} have enormously improved our understanding of the dynamics of photoexcited QPs on short
timescales, and the response of the superconducting state to weak pulsed laser excitation in
cuprates can now be unambiguosly identified on the femtosecond timescale \cite
{MgB2,nasiRT,NasiPRB1999,nasLSCO,Gedik,CaYBCO,YBCO124,kaindl,Schneider,Bianchi}.
Rothwarf and Taylor originally  proposed their phenomenological model for QP recombination in the framework of phonon-mediated pairing\cite{RT,nasiRT},  but spin-fluctuation mediated recombination  (which is relevant for the cuprates) is not excluded by their model. Both high-energy phonons and spin excitations could in principle mediate QP recombination, and the non-equilibrium studies so far did not directly reveal the pairing boson.

In this work we specifically address the question of the mediating boson by carefully measuring and analyzing the energy and time needed to destroy (vaporize) the superconducting condensate in La$_{2-x}$Sr$_{x}$CuO$_{4}$  with $x=0.1$ and 0.15.
We compare the measured vaporization energy with the thermodynamically measured condensation energy, and find that a substantial amount of energy is  temporarily stored by the glue boson bath during the vaporization process. Since spin and lattice subsystems have vastly different heat capacities, this places significant constraints on the type of bosonic bath which can mediate pairing.
By carefully considering the energy relaxation pathways
associated with pair-breaking dynamics we are able to conclude which bosonic excitations are involved in the destruction of the condensate, shedding light on the pairing boson responsible for superconductivity in these materials.

Outlining the sequence of events in our experiments microscopically, the
laser pump pulses first excite electrons from occupied to unoccupied states
within 1.5 eV of the Fermi level. Immediately afterwards, in an avalanche QP
multiplication process which
is well-studied in metals, as well as cuprates, the photoexcited carriers
relax to states near the Fermi energy via $\mathit{intra}$band electron-electron scattering, occuring on a typical timescale
$\tau _{e-e}\leq 50$ fs \cite{Perfetti}, and scattering with
phonons - preferentially interacting with
those phonons which are most strongly coupled to the QPs \cite%
{Allen} - resulting in significant non-equilibrium QP and phonon
populations within $\sim$100 fs of photoexcitation. The next relaxation step, QP recombination across a superconducting energy
gap (or pseudogap) with the emission of a boson with energy $\ge2\Delta$, takes significantly longer, and  is typically described very well by
the Rothwarf and Taylor model \cite{RT,nasiRT,NasiPRB1999}. The model does not directly identify the pairing boson, but two crucial parameters in the model do
depend on the electron-boson interaction, namely the characteristic bare
pair-breaking rate $\eta $ and recombination rate $R$ which define the pair-breaking and QP
recombination timescales respectively. \cite{RT,MgB2,nasiRT} Importantly, $\eta $ and $R$ can be determined
from the vaporization time $\tau_r$.\cite{MgB2}

The important feature of pump-probe experiments is that the transient density $n_{p}$ of the photoexcited (PE) QPs accumulated
at the gap edge is probed in real time by a time-delayed excited state
absorption process\cite{YBCO124}. The transient change in
reflectivity $\Delta R(t),$ of the probe pulse (which for small $\Delta R$
is linearly proportional to the transient photoinduced absorption \cite%
{comment}) is thus directly proportional to $n_{p}\left( t\right) $. This makes it possible to detect when the superconducting condensate is destroyed. \cite%
{NasiPRB1999,YBCO124}

In the experiments described here, we perturbed the superconducting
condensate in La$_{2-x}$Sr$_{x}$CuO$_{4}$ with 50 femtosecond laser pulses.
The experiments were performed on freshly cleaved surfaces of high quality La%
$_{2-x}$Sr$_{x}$CuO$_{4}$ ($x=0.1$ and $0.15$) single crystals with $%
T_{c}=30~$K and $38$ K respectively \cite{Sugai}. The laser pulses were
linearly polarized and incident along the $c$ axis of the crystal with a
wavelength of $\lambda =810$ nm ($\sim $1.5 eV). We used a Ti:Sapphire
oscillator and a 250 kHz amplifier to cover the range of
excitation fluences from $\mathcal{F}\sim 4\times 10^{-2}$ $\mu $J/cm$^{2}$
to $100 \mu$J/cm$^{2}$. The pump and probe beam diameters were measured
accurately with a pinhole and the absorbed energy  density was accurately determined\cite{Supp}. The low laser repetition rate of our laser ensured that there was no heat buildup between pulses even with the highest fluences used, and the temperature increase due to the laser was found to be less than 2 K (which can also be seen from a comparison of the $T_{c}$ measured optically with the $T_{c}$ from susceptibility measurements).

The photoinduced reflectivity change $\Delta R/R$ as a function of time
delay for different $\mathcal{F}$ is shown below $T_c$ ($T=4.5$ K) in Fig. 1a)
and above $T_c$ ($T= 32$ K) in Fig. 1b) for $x=0.1$ (the data for $x=0.15$ is
qualitatively the same). Below $T_{c}$ (Fig. 1a)) we
identify two relaxation processes with very different dynamics, which we
label as \textsf{A }and \textsf{B}. Signal
\textsf{B} is present from low $T$ to well above $T_{c}$ (Fig. 1b), and disappears gradually above the so called
pseudogap temperature $T^{\ast }$. In agreement with many previous low-$%
\mathcal{F}$ experiments \cite{NasiPRB1999,nasLSCO,CaYBCO,YBCO124}, it is
assigned to the carriers recombining across the pseudogap. Signal \textsf{A} is visible strictly only below $%
T_{c}$ and is - in accordance with previous works \cite{nasLSCO,Bianchi}- assigned to QP
recombination across the superconducting gap $\Delta _{s}(T)$, and has a relaxation time typically $\tau_A > $ 10ps at 4.5K  \cite{nasLSCO}. The rise-time $\tau_r =0.8\pm0.15$ ps of the superconducting signal $\Delta R/R_{A}$ is the
time required for the QP population to build up \cite{MgB2,nasiRT} by pair-breaking from the condensate.

Examining Figure 1 in more detail, we see that at low $%
\mathcal{F}$ and $T$ signal \textsf{A} is dominant. As fluence is increased,
the amplitude of signal \textsf{A}\ first increases with $\mathcal{F%
}$ and then starts to saturate for $%
\mathcal{F}$ above $\approx 12$ $\mu $J/cm$^{2}$. As signal \textsf{%
A} starts to saturate, signal \textsf{B}\ starts to become more visible, and
above\ the saturation threshold of signal A, a linear increase of the amplitude
of signal \textsf{B} with increasing $\mathcal{F}$ becomes clearly apparent.

The maximum amplitudes of $\Delta R_{\mathsf{A}}/R$ and $\Delta R_{\mathsf{B}}/R$
are shown in Fig. 2a) for $x=0.1$ and 0.15 as a function of $\mathcal{F}
$. We see that $\Delta R_{\mathsf{A}}/R$ is linear at low fluence for $\mathcal{F}<$ $8\mu $%
J/cm$^{2}$. Above $8\mu $
J/cm$^{2}$, the signal amplitude departs from linearity, indicating an onset of saturation associated with vaporization of the condensate.
$\Delta R_{\mathsf{A}}/R$ soon saturates and becomes constant for $\mathcal{F}>18\mu $J/cm$^{2}$ (up to the highest fluences measured).
In contrast, $\Delta R_{\mathsf{B}}/R$ is linear
with $\mathcal{F}$ both below and above $T_{c}$.

To accurately determine the
vaporization threshold, we carefully take into account the optical penetration depth $\lambda _{op}$ for the pump and the probe
beams and their spatial profile \cite{Supp}. From fits of the measured dependence of $%
\Delta R_{\mathsf{A}}/R$ on $\mathcal{F}$ to the function provided by a straightforward
model calculation (shown in Fig. 2a)), \cite{Supp} we obtain values for the threshold
vaporization fluence at 4.5 K: $\mathcal{F}_{T}=4.2\pm 1.7\mu \text{J/cm}^{2}$ for $%
x=0.1$ and $\mathcal{F}_{T}=5.8\pm 2.3\mu \text{J/cm}^{2}$ for $x=0.15$.

\begin{figure}[h]
\begin{center}
\includegraphics[width=7cm]{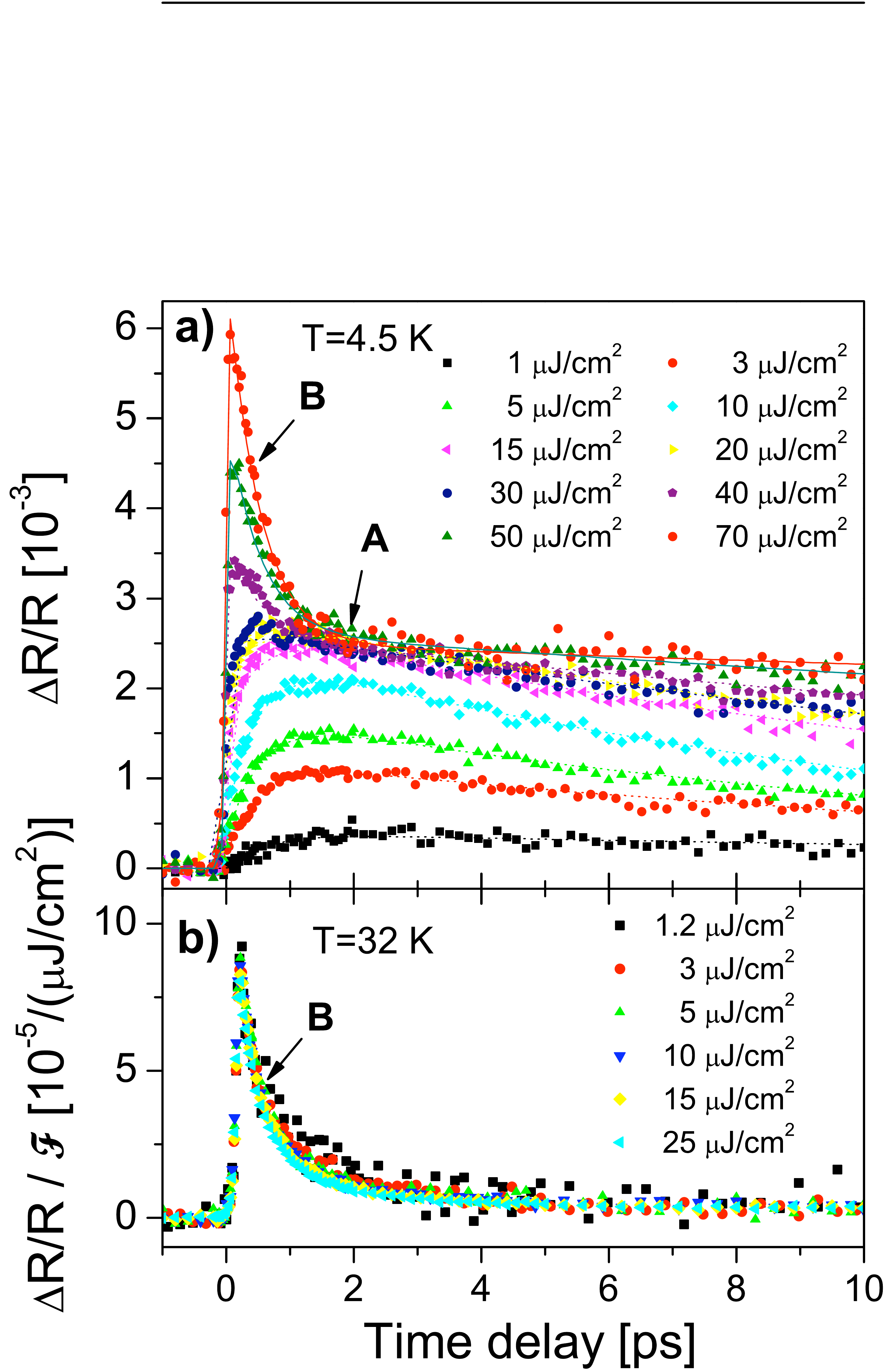}
\end{center}
\caption{(Color online).The photoinduced reflectivity $\Delta R/R$ in La$_{1.9}$Sr$_{0.1}$%
CuO$_{4}$ ($T_{c}$ = 30 K) taken at various photoexcitation fluences a) below
and b) above $T_{c}$. The data above $T_{c}$ are normalized with respect to $\mathcal{F}$ and fully overlap, showing that the response is linear. Below $T_c$ the two distinct relaxation components are marked as
\textsf{A} and \textsf{B}.}
\end{figure}

In Fig. 2b) we plot the $T$- dependence of $\Delta R_{\mathsf{A}%
}/R$ for several excitation levels for the $x=0.1$ sample. As expected, for $%
\mathcal{F}>\mathcal{F}_{T}$, the $T$-dependence of $\Delta R_{\mathsf{A}}/R$
does not depend on $\mathcal{F}$, since full vaporization is achieved at all
$T<T_c$. Near the threshold, for $\mathcal{F}=7\mu $J/cm$^{2}$, only partial
vaporization is evident and the amplitude $\Delta R_{\mathsf{A}}/R$ merges
with the high fluence data only as $T\rightarrow T_{c}$. We can understand
the T-dependence of the $\Delta R_{\mathsf{A}}/R$ by considering the
difference in reflectivity between the superconducting state and the normal state. The
induced change in reflectivity for fluences $\mathit{above}$ the
vaporization threshold $\emph{A}_{s}=\left\vert \frac{\Delta R}{R}%
\right\vert _{\mathcal{F}>\mathcal{F}_{T}}$ is proportional to $\sigma
_{1}^{n}-\sigma _{1}^{s}$, where $\sigma _{1}^{n}$ and $\sigma _{1}^{s}$ are
real parts of the complex conductivity in the normal and superconducting
states, respectively. Using the Mattis-Bardeen formulae \cite{MattisBardeen}
it follows that \cite{Supp}:

\begin{equation}
\emph{A}_{s}(T)\varpropto \frac{2\Delta (T)}{\hbar \omega }\ln \left( \frac{%
1.47\hbar \omega }{\Delta (T)}\right) \text{ \ ,}  \label{SatTdep}
\end{equation}%
where $\hbar \omega $ is the photon energy and $\Delta (T)$ the $T$%
-dependent gap. Using $\Delta (T)=\Delta _{0}\left( 1-\left( T/T_{c}\right)
^{2}\right) $ ($\Delta _{0}$ is gap at 0 K), which was previously found to describe $\Delta (T)$ in
cuprate superconductors \cite{Bozovic}, a very good agreement between Eq.(%
\ref{SatTdep}) and the data for $\mathcal{F}>\mathcal{F}_{T}$ is obtained
(see Figure 2 \textsf{b}).

\begin{figure}[h]
\begin{center}
\includegraphics[width=8.5cm]{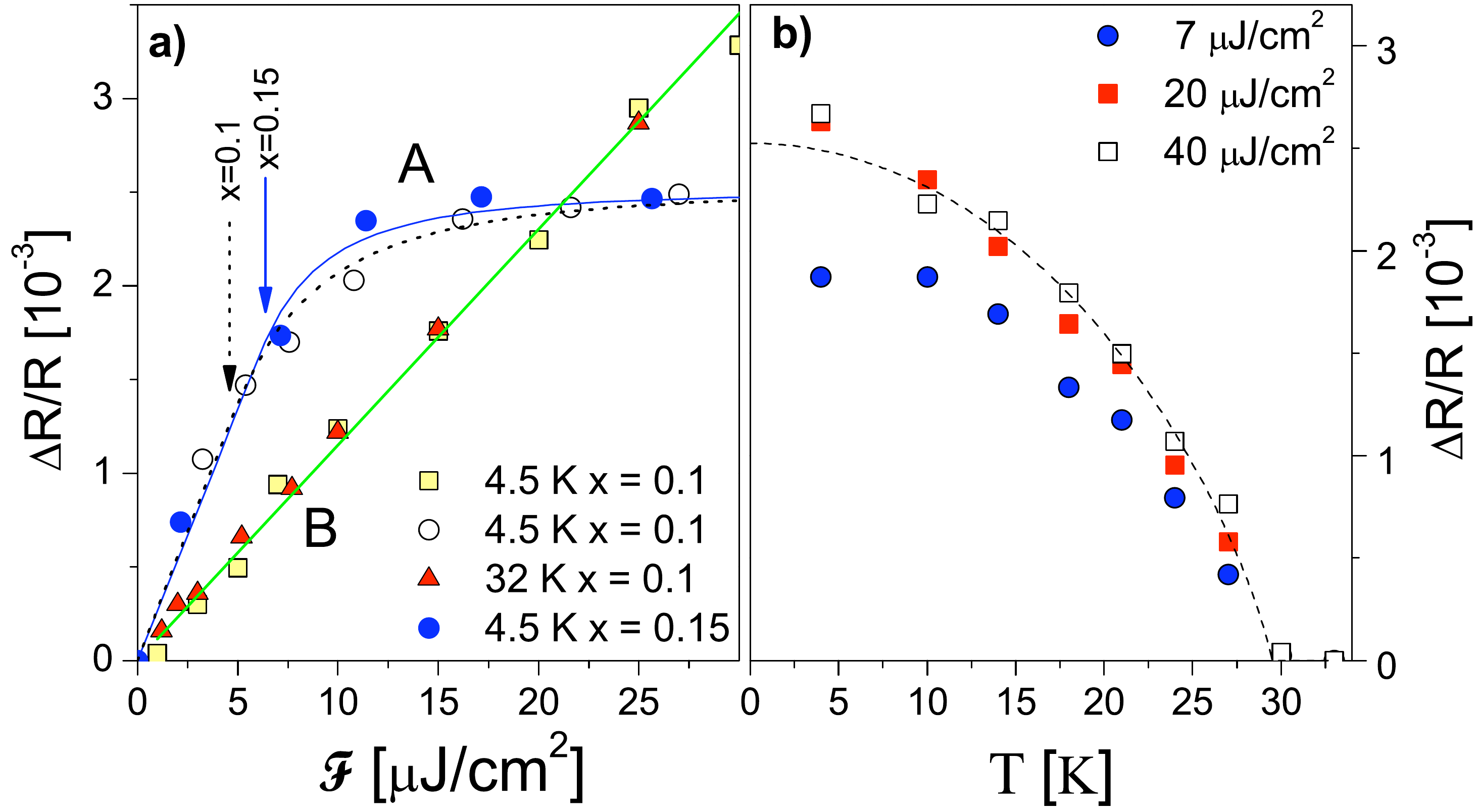}
\end{center}
\caption{(Color online). a)\ The maximum amplitude  $\Delta R_{\mathsf{A}}/R$ at 4.5 K for $x=0.1$ (empty circles) and $x=0.15$ (full circles), and $\Delta R_{\mathsf{B}}/R$ and at 4.5 K (squares) and 32 K (triangles) for x=0.1 in La$_{2-x}$Sr$_x$CuO$_4$.
The arrows mark the vaporization thresholds $\mathcal{F}_{T}=4.2\pm 1.7\mu \text{J/cm}^{2}$ and $\mathcal{F}%
_{T}=5.8\pm 2.3\mu\text{J/cm}^{2}$ for x=0.1 and x=0.15 respectively
obtained from the fit\cite{Supp} (lines). b) The $T$-dependence of $\Delta R_{A}/R$ for x=0.1. The dashed line is a fit to the data using Eq.(1).}
\end{figure}

Let us now examine the energy relaxation pathways on the pair-breaking
 timescale of $\sim1$ ps. Phonons released during
this time need at least $\lambda _{op}/v_{s}$ $\sim 30$ ps to escape
from the excited volume, $v_{s}$ being the velocity of sound. The
characteristic QP diffusion time from the excitation volume is also of the
order of $\sim $ 100 ps, calculated using the measured QP diffusion constant
for very clean samples of YBa$_{2}$Cu$_{3}$O$_{6.5}$ at 4 K \cite{Orenstein}. Therefore we can conclude that the absorbed optical pulse
energy cannot diffuse or escape, and remains in the excitation volume on the timescale of 1 picosecond.


Next, let us analyze the microscopic energy relaxation processes within the excitation volume in more detail.
The energy densities in the excitation volume at vaporization threshold for $x=0.1$ and $x=0.15$ shown in Fig.2 a) are $U_{p} = \mathcal{F}_{T}/(\lambda _{op}k_{B})=2.0\pm 0.8$ K/Cu and $%
2.6\pm 1.0$ K/Cu respectively (using
$\lambda _{op}=150$ nm at 810 nm \cite{OpticalProp})). Both are significantly higher than the thermodynamically
measured condensation energies extracted from specific heat data, which are  $U_{c}/k_{B}=0.12$ K/Cu
for $x=0.1$, and $U_{c}/k_{B}=0.3$ K/Cu for $x=0.15$
\cite{Matsuzaki}. The ratio of the two energies are thus
$U_{p}/U_{c}\simeq 16$ and $8.5$ respectively. This means that a significant
amount of energy $(U_{p}-U_{c})$ is not directly used in the vaporization
process, but is stored elsewhere on the timescale of $\tau_r$.

There are excitations of the system, such as phonons of different symmetry,
but also potentially spin fluctuations, etc., that make up the difference
between the condensation energy and the measured optical vaporization energy. Let us
consider spin excitations first. The energy required to heat the entire spin
bath from 4.5 K to $T_{c}$ for $x=0.1$ is given by $U_{M}=\int
\nolimits_{4.5K}^{T_{c}=30K}C_{M}(T)dT$, where $C_{M}(T)$ is the magnetic
specific heat. Using the published value\cite{CM} of $C_{M}(T)$ for undoped
La$_{2}$CuO$_{4}$ ($C_{M}$ in doped La$_{1.9}$Sr$_{0.1}$CuO$_{4}$ can only
be smaller), we obtain $U_{M}\simeq 80$ mJ/mol (0.01 K/Cu). Clearly, the magnetic
system alone is not capable of absorbing ($U_{p}-U_{c})/k_{B}\approx 1.9$
K/Cu, its heat capacity being too small by a factor of $\sim 190$. Making the
same estimate for the lattice excitations, we obtain $U_{L}=\int
\nolimits_{4.5K}^{T_{c}}C_{p}(T)dT$ $\simeq 77$ J/mol (9 K/Cu) for $x=0.1$ ($T_{c}=30$ K)
and 240  J/mol (28 K/Cu) for $x=0.15$ ($T_{c}=38$ K), where $C_{p}(T)$ is the
experimentally measured specific heat\cite{Matsuzaki}. The phonon
subsystem can thus easily absorb the excess supplied energy, with $%
U_{L}/U_{p}\sim 4.5$ for $x=0.1$ (and 11.6 for $x=0.15$).


This observed discrepancy between measured $U_{p}$ and thermodynamically
measured condensation energy, as well as the $T$- and
$\mathcal{F}$-dependence of the superconducting state depletion process can
be naturally explained within the Rothwarf-Taylor (RT) model in the
bottleneck regime, where the pairing bosons are reaching quasi-equilibrium with the QPs \cite{nasiRT} on the 1 ps timescale and share some of the energy supplied by the optical pulses.

In the RT model, the pair-breaking time (which corresponds to the condensate vaporization time when $\mathcal{F}>\mathcal{F}_T$) is given by $\tau
_{r}^{-1}=\eta \sqrt{1/4+(4N(0)+2n(0))R/\eta }$ where the initial QP and boson densities are $n(0)$ and $N(0)$ respectively \cite{nasiRT,MgB2}.  For weak
photoexcitation, when both $n(0)$, $N(0)< n_T$, where the threshold density is defined as
$n_T=\eta /R$, $\tau_{r}$ is independent of $\mathcal{F}$, and $2\tau_{r}=\eta ^{-1}$.
For intense photoexcitation, when either $n(0)$, $N(0)\gtrsim n_T$, $\tau_{r}$ $strongly$ depends on $\mathcal{F}$.
A strong $%
\mathcal{F}$-dependence of $\tau_{r}$ is $not$ observed in our data, which implies that LSCO is in the "weak"
perturbation regime over our range of $\mathcal{F}$, and so $\eta = 1/(2 \tau_r) \approx 0.5\times10^{12} s^{-1}$. To estimate $n_{T}$, we take  $R=0.1$ cm$^{2}s^{-1}$ measured by Gedik \textit{et al.}
in YBCO \cite{Gedik} and
obtain a threshold density $n_{T}=\eta /R\approx 0.8\times 10^{20}$cm$^{-3}\approx 0.8\times
10^{-2}$/Cu.
We can make an alternative microscopic estimate of $n_T$ using the formula for the bare recombination rate from ref. \cite{Hg} (with phonons as the mediating bosons) $R=\frac{8\pi \Lambda \Delta ^{2}}{\hbar ^{3}\Omega
_{D}^{2}N_{0}}$, where $N_{0}$ is the density of states at the
Fermi level, $\Delta $ is the superconducting gap, $\Omega _{D}$ is the
characteristic phonon frequency and $\Lambda$ the electron-phonon coupling
constant (which is the same as appears in the McMillan formula for $T_c$ \cite{Hg}). Taking typical values for LSCO $N_{0}=5/eVCu$, $\Delta \approx
0.01eV$ , $\Omega _{D}\approx 0.1eV$ and the measured $\Lambda =0.9$ \cite{chekalin}, we
obtain $R=0.7\times 10^{-8}$ cm$^{3}$s$^{-1}$ which gives a very similar threshold density as the phenomenological estimate $n_{T}=\eta /R=1.5\times
10^{-2}$/Cu.  Note that both are just slightly lower than the estimated photoexcited QP density at threshold fluence which is $n_{p}^{s}=\frac
{\mathcal{F}}{\Delta_{s}\lambda_{op}}\frac{\text{e}-1}{\text{e}}\simeq2.7\times10^{-2}$ /Cu.
We can conclude that the RT model involving
phonons in the pair-breaking process gives a self-consistent
quantitative description of the vaporization dynamics.

Let us now see whether the relaxation processes on the sub- 1 ps
timescale might somehow involve spin excitations. In this scenario, energy might be initally
transferred from PE carriers to the spin subsystem on a
timescale much shorter than 1 ps and QPs would then be excited from the condensate by absorbing
energy from the hot bath of spin excitations. For energy relaxation only real (not virtual \cite{Sugai}) processes are relevant and the relevant
interaction between QPs and spin excitations is spin-orbit coupling. Such a
scenario is consistent with our data, provided that the spin-orbit
relaxation time $\tau _{S-O}$ is equal to, or shorter than the observed vaporization time
of $\tau _{r}=0.8\pm0.15$ ps. To estimate the vaporization time for this case, we use the fact that spin-lattice relaxation is a
process in which electron-phonon relaxation follows spin-orbit relaxation, and
 $\tau _{S-L}\simeq \tau _{S-O}+\tau _{E-P}$. So, for spin
excitations to be involved in the pair breaking and QP relaxation process, $\tau _{S-L}$ needs
to be of the order of 1 ps or less. Electron-paramagnetic resonance (EPR) measurements of Cu spin relaxation in La$_{1.9}$Sr$_{0.1}$%
CuO$_{{4}}$, give EPR linewidths ranging from $\Delta H\sim 1$
kilogauss at 30 K to $\Delta H=$ $3$ kilogauss at 8 K. This corresponds to a lower limit of the
relaxation time $\tau _{S-L}\simeq 100-340$ ps \cite{Kochelaev2}, which is
much longer than observed.  Assuming that the measured $\tau _{S-L}$ is correct, the pair-breaking thus cannot proceed via the spin
excitations, because the relaxation process at 4.5 K would take over $ 340$ ps,
instead of $\sim 0.8$ ps. Thus spin excitations cannot be responsible for the destruction of the
superconducting condensate by any currently known spin-orbit relaxation mechanism.
This conclusion has important implications for the pairing mechanism in these compounds. The pair-breaking process discussed above is related to QP recombination (pairing) by time-reversal symmetry, and therefore both processes must involve the same mediating boson, i.e. phonons.
We conclude that only phonon-mediated vaporization is consistent
with the observed dynamics, effectively ruling out spin-mediated QP recombination and pairing in these materials.

We wish to acknowledge valuable discussions and important comments from K.
Alex Muller, N. Ashcroft, P. B. Allen, A. S. Alexandrov, D. Van der Marel,
E. Maksimov, I. Bozovic and D. Newns.

\newpage

\section{Supplementary material}

\subsection{Calculation of the behavior of the photoinduced reflectance as a function of fluence $\mathcal{F}$ for a superconductor in the high density regime}

To accurately identify the point where the superconducting condensate is vaporized, we need to account for geometrical aspects due to the finite absorbtion length of pump and probe light as well as the transverse beam profiles.

In the probe beam, the sample penetration depth has to be accounted for twice
(upon entering and exiting the sample). The relative photoinduced change in reflectivity is then given by:

\begin{equation}
\frac{\Delta R}{R} \propto \int_{0}^{\infty }\int_{0}^{\infty }
{\exp {\left( -2\frac{z}{\lambda _{op}}\right) }
{\exp {\left( -\frac{r^{2}}{\rho _{pr}^{2}}\right)} n_{qp}(r,z)rdr}dz}
\end{equation}

\noindent where $\lambda _{op}$ is optical penetration depth, and $2\rho
_{pr}$ is probe beam diameter on the sample.
We take the change of
reflectivity to be linear with the density of photoexcited quasiparticles $n_{qp}$, and $n_{qp}$
to be approximately linear with excitation density up to the threshold excitation fluence $\mathcal{F}_{T}$ where the superconducting condensate is evaporated. For $\mathcal{F}>\mathcal{F}_{T}$ all quasiparticles are excited and $n_{QP}$ saturates at $n_{s}$

\begin{equation}
n_{qp} \thickapprox \left\{
\begin{array}{lc}
\frac{\mathcal{F}(r,z)}{\mathcal{F}_{T}}n_{s} & \mathcal{F}(r,z)<\mathcal{F}_{T}\\
n_{s} & \mathcal{F}(r,z)>\mathcal{F}_{T}\end{array}
\right.
\end{equation}%
The light fluence penetrating into the sample is $\mathcal{F}_{0} = (1-R)\mathcal{F}_{pu}$ where $\mathcal{F}_{pu}$ represents the laser fluence on the surface of the sample. The laser fluence within sample is:
\begin{equation}
\mathcal{F}(r,z)=\mathcal{F}_{0}\exp {\left( -\frac{z}{\lambda _{op}} \right) }\exp {\left( -\frac{r^{2}}{\rho _{pu}^{2}}\right) }
\end{equation}

Assuming that in the normal state $\Delta R/R$ does not depend on $\mathcal{F}$  due to other processes, we calculate the integral in two parts: for $\mathcal{F}(r,z) < \mathcal{F}_{T}$ and separately over the volume where $\mathcal{F}_{T} > \mathcal{F}_{T}$. With pump beam radius $\rho_{pu}$ and $\frac{1}{\rho _{eff}^{2}}=\left( \frac{1}{\rho_{pu}^{2}}+\frac{1}{\rho _{pr}^{2}}\right)$ the complete integral is:

\begin{equation*}
\frac{\mathcal{F}_{0}}{\mathcal{F}_{T}}n_{s}\!\!\int_{0}^{\infty }\!\!\!{\exp {\left( -\frac{3z}{
\lambda_{op}}\right) }\left[ \int_{0}^{\infty }\!\!\!{\exp {\left( \frac{-r^{2}}{%
\rho _{eff}^{2}}\right) }\Theta \left(1- \frac{\mathcal{F}(r,z)}{\mathcal{F}_{T} } \right)} rdr\right] dz}
\end{equation*}
\begin{equation}
+n_{s}\int_{0}^{\infty }\!\!\!{\exp {\left( -\frac{2z}{\lambda
_{op}}\right) }\left[ \int_{0}^{\infty }\!\!\!{\exp {\left( \frac{-r^{2}}{%
\rho_{pr}^{2}}\right) }
\Theta \left( \frac{\mathcal{F}(r,z)}{\mathcal{F}_{s} }-1 \right) rdr}\right] dz}
\end{equation}
where $\Theta$ is Heaviside step function. The behavior of this integral depends on the ratio $ \frac{\mathcal{F}_{0}}{\mathcal{F}_{T}}$. To get a dimensionless result we normalize the result with its saturated value:
\begin{equation}
\frac{\Delta R_{s}}{R} \propto n_{s}\int_{0}^{\infty }\int_{0}^{\infty } {\exp {\left( -2\frac{z}{\lambda_{op}}\right) }
{\exp {\left( -\frac{r^{2}}{\rho _{pr}^{2}}\right)} rdr}dz} = n_{s}\frac{\rho_{pr}^{2}\lambda _{op}}{4}
\end{equation}
For $\mathcal{F}_{0}<\mathcal{F}_{T}$ the dependence is linear in $\mathcal{F}$:
\begin{equation}
\frac{\Delta R}{\Delta R_{s}} = \frac{2\mathcal{F}_{0}\rho _{eff}^{2}}{3\mathcal{F}_{T}\rho_{pr}^{2}}
= \frac{2}{3}f^{-1}(1+\bar{\rho}^{-2})
\label{EgGeomCorrLin}
\end{equation}
while for $\mathcal{F}_{0}>\mathcal{F}_{T}$ :%

\begin{equation}\
 \frac{\Delta R}{\Delta R_{s}} =
    \left[
         1- f^{2} +
        \frac{2 \left( f^{\bar{\rho}^{2}}  - f^{2}\right)  }{\bar{\rho}^{2} - 2} +
         \frac{6 f^{\bar{\rho}^{2}} /(1+\bar{\rho}^{-2})- 2f^{2}\bar{\rho}^{2}}
         {6-3\bar{\rho}^{2}}
     \right],
\label{EgGeomCorrNonLin}
\end{equation}
with $\bar{\rho}^{2} = \rho_{pu}^{2}/\rho_{pr}^{2}$ and $f=\frac{\mathcal{F}_{T}}{\mathcal{F}_{0}}$. As can be seen from eq. \ref{EgGeomCorrLin} one can determine the saturated value of excitation fluence $\mathcal{F}_{T}$ simply by reading the value at which $\Delta R/R$ reaches $2\rho _{eff}^{2}/3\rho_{pr}^{2}$ of the maximum (saturation) value of $\Delta R/R$.

To determine $\mathcal{F}_{pu}$ in our measurements we used a pinhole of diameter $2r_{pin}$ and measured the power of the beam in front of the pinhole ($P_{in}$) and after the pinhole ($P_{tr}$). With known repetition rate of the pulses ($\nu_{rr}$) we can calculate $\mathcal{F}_{pu}$ and diameter of the beam $\rho$:
\begin{equation}
   P_{tr}/P_{in} = \frac{\mathcal{F}_{pu}\nu_{rr}\int_{0}^{r_{pin} }{\exp {\left( -\frac{r^{2}}{\rho^{2}}\right)} 2\pi rdr}}{\mathcal{F}_{pu}\nu_{rr}\int_{0}^{\infty }{\exp {\left( -\frac{r^{2}}{\rho^{2}}\right)} 2\pi rdr}}
\end{equation}
and
\begin{equation}
P_{in} = \mathcal{F}_{pu}\nu_{rr}\int_{0}^{\infty }{\exp {\left( -\frac{r^{2}}{\rho^{2}}\right)} 2\pi rdr} = \mathcal{F}_{beam}\pi \rho^{2}\nu_{rr}
\end{equation}
To obtain the $\mathcal{F}_{0}$ we accounted for the light reflected from the cryostat window (8\%) and the part reflected from the sample ($R$).

The excitation energy density calculated from the absorbed energy fluence is:
\begin{equation}
    U =  \frac{ \mathcal{F}(r,z)}{\lambda_{op}}
\end{equation}

\begin{figure}[t]
\begin{center}
\includegraphics[width=9.0cm]{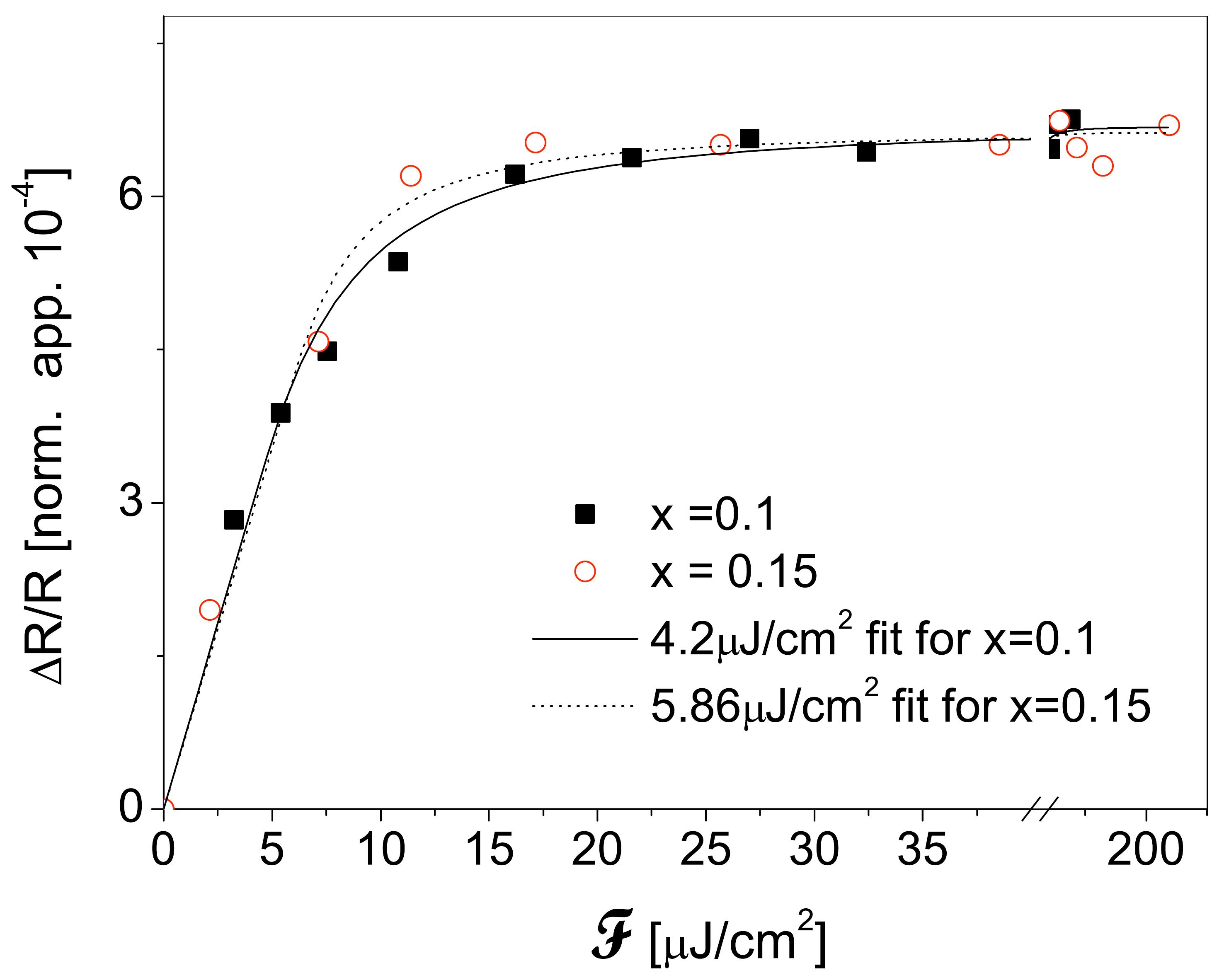}\\[0pt]
\end{center}
\caption{Amplitude as a function of fluence. The $\rho_{pu}:\rho_{pr}$ ratio is 1.5 for sample $x = 0.1$ and 4 for $x = 0.15$. The lines are best fits of the data to eq. \ref{EgGeomCorrLin} and \ref{EgGeomCorrNonLin}.}
\label{FigGeomCorrection}
\end{figure}

In our experiment we used the following parameters. The pump/probe beams radii were (42/28 ($\pm15\%$)$\mu$m) and (71/18 ($\pm15\%$) $\mu$m) for the samples $x=0.1$ and 0.15 respectively. $\mathcal{F}_{0}$ was determined with the relative accuracy of 0.3. Penetration depths were calculated from the optical conductivity and the dielectric function data \cite{OpticalProp} and were found to be 147$\pm15$ nm for $x=0.1$ and 156$\pm15$ nm for $x=0.15$. The reflectivity is 0.135 for $x=0.1$ and 0.130 for $x=0.15$ \cite{OpticalProp}.

The saturated fluences obtained by fitting the eqs. 6 and 7 to the data (see Fig. \ref{FigGeomCorrection}) are $\mathcal{F}_{T}^{0.1} = $4.2$ \pm 1.7\mu$Jcm$^{-2}$ and $\mathcal{F}_{T}^{0.15} = 5.86\pm2.3 \mu$Jcm$^{-2}$ for underdoped (x=0.1) and optimally doped (x=0.15) samples respectively. This is equivalent to the absorbed energy of  1.95$\pm$0.78$k_{B}$K per Cu atom for x=0.1 and 2.56$\pm$1.0$k_{B}$K for x=0.15.

\subsection{The temperature dependence of the photoinduced change in
reflectivity in the limit of condensate vaporization.}

In the following we describe the derivation of the expected temperature
dependence of the photoinduced change in reflectivity at optical frequencies
in the limit of excitation intensities higher than the condensate vaporization treshold.%

Figure 2a) of the main text shows the time-variation of the photoinduced reflectivity trace recorded
at 4.5 K on La$_{1.9}$Sr$_{0.1}$CuO$_{4}$ ($T_{c}$=30K) as a function of the
photoexcitation fluence ranging from 1 to 200 $\mu$J$/$cm$^{2}$. Component
\textsf{A}, which describes the dynamics of the superconducting state
pair-breaking and recovery, exhibits clear saturation at high excitation
intensities. Analysis of the $\mathcal{F}$ -dependence of the signal amplitude at $t=2$
ps with the model described in the previous section gives the treshold fluence $\mathcal{F}_{T}$ $=$4.2$\mu$J$/$cm$^{2}$ for x=0.1 and  $\mathcal{F}_{T}$ $=$5.9$\mu$J$/$cm$^{2}$ for x=0.15.
It follows from these data that at $\mathcal{F}>\mathcal{F}_{T}$ the
superconducting state is vaporized on the timescale of $\approx1$ ps after
excitation with 50 fs optical pulses.

The temperature dependence of the induced change in reflectivity at
$\mathcal{F}>\mathcal{F}_{T}$ is markedly different than the temperature
dependence obtained in the low excitation regime, which is not surprising.
Indeed, the amplitude of the induced change in reflectivity in the regime of
condensate vaporization, \emph{A}$_{s}$, is given by
\begin{equation}
\emph{A}_{s}=\left\vert \frac{\Delta R}{R}\right\vert _{\mathcal{F}%
>\mathcal{F}_{T}}=\frac{R_{n}-R_{s}}{R_{s}}\simeq\frac{R_{n}-R_{s}}{R_{n}%
}\text{\ \ \ \ \ ,}%
\end{equation}
where $R_{n}$ and $R_{s}$ are the reflectivities in the normal and in the
superconducting states, respectivey, and $R_{n},R_{s}\gg R_{n}-R_{s}$. At
optical frequencies, the induced change in reflectivity is proportional to the
induced change in the imaginary component of the refraction index and
therefore proportional to the induced change in the real part of the optical
conductivity $\frac{\Delta R}{R}\varpropto\frac{\Delta k}{k}\varpropto
\frac{\Delta\varepsilon_{2}}{\varepsilon_{2}}\varpropto\frac{\Delta\sigma_{1}%
}{\sigma_{1}}$ giving%

\begin{equation}
\emph{A}_{s}=\left\vert \frac{\Delta R}{R}\right\vert _{\mathcal{F}%
>\mathcal{F}_{T}}\varpropto\frac{\sigma_{1}^{n}-\sigma_{1}^{s}}{\sigma_{1}%
^{n}}\text{ \ \ \ .}%
\end{equation}
To determine the temperature dependence of $\emph{A}_{s}$ we evaluate
$\frac{\sigma_{1}^{s}}{\sigma_{1}^{n}}(T)$ where $\frac{\sigma_{1}^{s}}%
{\sigma_{1}^{n}}$ is given by the Mattis-Bardeen relation\cite{MattisBardeen}
\begin{equation}
\frac{\sigma_{1}^{s}}{\sigma_{1}^{n}}\left(  \hbar\omega\right)  =\frac
{2}{\hbar\omega}%
{\displaystyle\int\nolimits_{\Delta}^{\infty}}
\left(  f\left(  \varepsilon\right)  -f\left(  \varepsilon+\hbar\omega\right)
\right)  g\left(  \varepsilon\right)  d\varepsilon+\frac{1}{\hbar\omega}%
{\displaystyle\int\nolimits_{\Delta-\hbar\omega}^{-\Delta}}
\left(  1-2f\left(  \varepsilon+\hbar\omega\right)  \right)  g\left(
\varepsilon\right)  d\varepsilon\text{ \ \ .}\label{MatBarEq}%
\end{equation}
Here $f\left(  \varepsilon\right)  $ is the Fermi-Dirac distribution function,
$\hbar\omega$ is the photon energy, $\Delta$ is the superconducting gap, and
$g\left(  \varepsilon\right)  $ is%

\begin{equation}
g\left(  \varepsilon\right)  =\frac{\varepsilon\left(  \varepsilon+\hbar
\omega\right)  +\Delta^{2}}{\sqrt{\varepsilon^{2}-\Delta^{2}}\sqrt{\left(
\varepsilon+\hbar\omega\right)  ^{2}-\Delta^{2}}}\text{ \ .}%
\end{equation}
In the limit of $\Delta\ll\hbar\omega$ \ Eq.(\ref{MatBarEq}) can be rewritten
as%
\begin{equation}
\frac{\sigma_{1}^{s}}{\sigma_{1}^{n}}\left(  \hbar\omega\right)  \simeq
\frac{1}{\hbar\omega}%
{\displaystyle\int\nolimits_{\Delta}^{\hbar\omega-\Delta}}
g\left(  \varepsilon-\hbar\omega\right)  d\varepsilon+\frac{2}{\hbar\omega}%
{\displaystyle\int\nolimits_{\Delta}^{\infty}}
f\left(  \varepsilon\right)  \left(  g\left(  \varepsilon\right)  -g\left(
\varepsilon-\hbar\omega\right)  \right)  d\varepsilon\text{ .}\label{I1I2}%
\end{equation}
The first term on the right hand side of Eq.(\ref{I1I2}) can be evaluated
exactly and is given by\cite{MattisBardeen}%

\begin{equation}
I_{1}=\left(  1+\frac{2\Delta}{\hbar\omega}\right)  E\left(  \frac
{1-\frac{2\Delta}{\hbar\omega}}{1+\frac{2\Delta}{\hbar\omega}}\right)
-\frac{4\Delta}{\hbar\omega}K\left(  \frac{1-\frac{2\Delta}{\hbar\omega}%
}{1+\frac{2\Delta}{\hbar\omega}}\right)  \text{ \ ,}%
\end{equation}
where $E$ and $K$ are the complete elliptic integrals. In the limit of photon
energies being much higher than the gap, $\frac{2\Delta}{\hbar\omega}\ll1$, it
then follows that%

\begin{equation}
I_{1}\simeq1-\frac{2\Delta}{\hbar\omega}\ln\left(  \frac{4\hbar\omega}%
{e\Delta}\right)  \text{ \ .}%
\end{equation}
The second term on the right hand side of Eq.(\ref{I1I2}) can also be
calculated exactly in the limit of $T,\Delta\ll\hbar\omega$ and is found to be%
\begin{equation}
I_{2}\simeq\frac{\sqrt{8\pi\Delta k_{B}T}}{\hbar\omega}\exp\left(
\frac{-\Delta}{k_{B}T}\right)  \text{ \ .}%
\end{equation}
Clearly, $I_{2}$ presents only a small correction which becomes noticable only
in the close vicinity to T$_{c}$. Therefore, $\emph{A}_{s}(T)$ is in the limit
when $\hbar\omega\gg T,\Delta$ given by
\begin{equation}
\emph{A}_{s}(T)\varpropto\frac{2\Delta(T)}{\hbar\omega}\ln\left(  \frac
{4\hbar\omega}{e\Delta(T)}\right)  .\label{SatTdep}%
\end{equation}
Due to the fact that the data on the
temperature dependence of the gap are fairly scarce, $\Delta(T)$ has been
commonly assumed to follow the mean-field (BCS-like) temperature dependence.
Comparison of $\emph{A}_{s}(T)$, where a BCS T-dependence of $\Delta(T)$ is
assumed, to the experimentally measured $\left\vert \frac{\Delta R}%
{R}\right\vert _{\mathcal{F}>\mathcal{F}_{T}}$ shows, however, that
$\Delta(T)$ has a substantially weaker T-dependence than BCS functional form.
In fact, from the available data on the temperature dependence of $\Delta(T)$
in cuprate superconductors, which is obtained from the temperature dependence
of the SIS tunneling junction caracteristics\cite{Bozovic}, as well as from
the intrinsic tunneling data\cite{Krasnov} it follows that the gap follows
$\Delta(T)=\Delta_{0}\left(  1-\left(  T/T_{c}\right)^{2}\right)  $
temperature dependence over a wide temperature range. Indeed, using this
functional form with $\hbar\omega=1.5$ eV, and $2\Delta_{0}=120$ , a nearly perfect agreement between $\emph{A}_{s}(T)$ and the
experimentally measured $\left\vert \frac{\Delta R}{R}\right\vert
_{\mathcal{F}>\mathcal{F}_{T}}$ is found (Figure 2).%
We should note however, that unlike in the low excitation regime, where by
fitting the temperature dependence of the amplitude of the photoinduced
change in reflectivity the magnitude of the superconducting gap can be
extracted, this is not the case in the high excitation regime.

\begin{figure}
[h]
\begin{center}
\includegraphics[width=9cm]{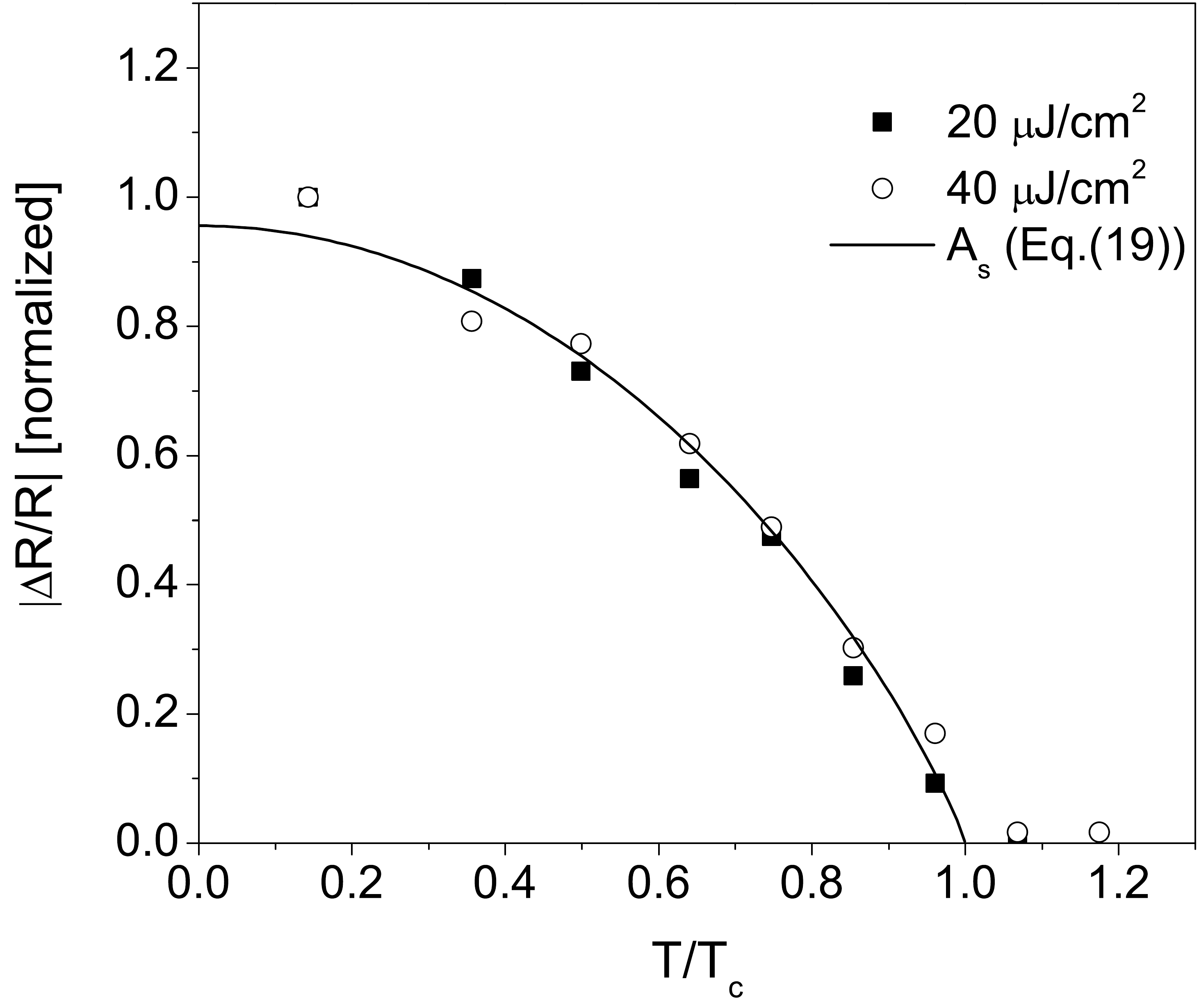}%
\caption{The temperature dependence of the experimentally measured $\left\vert
\frac{\Delta R}{R}\right\vert _{\mathcal{F}>\mathcal{F}_{T}}$ for
$\mathcal{F}=20$ and $40$ $\mu$J/cm$^{2}$, compared to $\emph{A}_{s}(T)$ given
by Eq.(\ref{SatTdep}) with $\Delta(T)=\Delta_{0}\left(  1-\left(
T/T_{c}\right)  ^{2}\right)  $ temperature dependence of the gap (solid line).
}%
\end{center}
\end{figure}


\begin{thebibliography}{99}
\bibitem{Testardi} L.R. Testardi, \textit{Phys. Rev.} B \textbf{4}, 2189
(1971).

\bibitem{Han} S.G. Han, Z.V. Vardeny, K.S. Wong, O.G. Symko, and G. Koren,
\textit{Phys. Rev. Lett.} \textbf{65}, 2708 (1990).

\bibitem{Chwalek} J.M. Chwalek, C. Uher, J.F. Whitaker, G.A. Mourou, and
J.A. Agostinelli, \textit{Appl. Phys. Lett.}, \textbf{58}, 980 (1991).

\bibitem{Albrecht} W. Albrecht, Th. Kruse, and H. Kurz, \textit{Phys. Rev.
Lett.} \textbf{69}, 1451 (1992).

\bibitem{Stevens} C.J. Stevens, \textit{et al.}, \textit{Phys. Rev. Lett.}
\textbf{78}, 2212 (1997).

\bibitem{RT} A. Rothwarf and B.N. Taylor, \textit{Phys. Rev. Lett.} \textbf{%
19}, 27 (1967).

\bibitem{MgB2} J. Demsar, \textit{et al.}, \textit{Phys. Rev. Lett.} \textbf{%
91}, 267002 (2003); J. Demsar, \textit{et al.}, \textit{Int. J. Mod. Phys.}
\textbf{B 17}, 3675 (2003).

\bibitem{nasiRT} V.V. Kabanov, J. Demsar, D. Mihailovic, \textit{Phys. Rev.
Lett.} \textbf{95}, 147002 (2005).

\bibitem{NasiPRB1999} V.V. Kabanov, J. Demsar, B. Podobnik, D. Mihailovic,
\textit{Phys. Rev. B} \textbf{59}, 1497 (1999).

\bibitem{Gedik} N. Gedik, \textit{et al.}, \textit{Phys. Rev. B} \textbf{70}%
, 014504 (2004), G. P. Segre, \textit{et al.}, \textit{Phys. Rev. Lett.}
\textbf{88}, 137001 (2002). R.A. Kaindl, \textit{et al.}, \textit{Phys. Rev.
B} \textbf{72}, 060510 (2005).

\bibitem{nasLSCO} P. Kusar, J. Demsar, D. Mihailovic, S. Sugai, \textit{%
Phys. Rev. B} \textbf{72}, 014544 (2005).

\bibitem{CaYBCO} J. Demsar, B. Podobnik, V.V. Kabanov, Th. Wolf, D.
Mihailovic, \textit{Phys. Rev. Lett.} \textbf{82}, 4918 (1999).

\bibitem{YBCO124} D. Dvorsek, \textit{et al.}, \textit{Phys. Rev. B},
\textbf{66}, 020510 (2002).

\bibitem{kaindl} R.A. Kaindl, \textit{et al.}, \textit{Science} \textbf{287}%
, 470 (2000).

\bibitem{Schneider} M.L. Schneider, \textit{et al.} \textit{Eurphys. Lett. }%
\textbf{60}, 460 (2002), ibid. \textit{Phys.Rev. B }\textbf{70}, 012504
(2004).

\bibitem{Bianchi} G. Bianchi, C. Chen, M. Nohara, H. Takagi, J.F. Ryan,
\textit{Phys. Rev. B} \textbf{72}, 094516 (2005).

\bibitem{Perfetti} L. Perfetti, et al., \textit{Phys. Rev. Lett., } \textbf{99}, 197001 (2007).

\bibitem{Allen} P.B. Allen, \textit{Phys. Rev. Lett.} \textbf{59} 1460
(1987).

\bibitem{comment} For larger fluences, we can expect deviations from
linearity due to photoinduced changes of the electronic band structure.

\bibitem{Sugai} S. Sugai, H. Suzuki, Y. Takayanagi, \ T. Hosokawa, N.
Hayamizu, \textit{Phys. Rev. B} \textbf{68}, 184504 (2003).

\bibitem{Supp} See Supplementary material.

\bibitem{MattisBardeen} D.C. Mattis, J. Bardeen, \textit{Phys. Rev.}\textbf{%
111}, 412 (1958).

\bibitem{Bozovic} I. Bozovic, J.N. Eckstein, \textit{Appl. Surf. Sci.}
\textbf{113/114}, 189 (1997), V.M. Krasnov, A. Yurgens, D. Winkler, P.
Delsing, T. Claeson, \textit{Phys. Rev. Lett.} \textbf{84}, 5860 (2000).

\bibitem{Orenstein} N. Gedik, J. Orenstein, R. Liang, D.A. Bonn, W.N. Hardy,
\textit{Science }\textbf{300}, 1410 (2003).

\bibitem{Matsuzaki} T. Matsuzaki, N. Momono, M. Oda, M. Ido, \textit{J.
Phys. Soc. Jpn.}, \textbf{73}, 2232 (2004).

\bibitem{CM} M.R. Singh, S.B. Barrie, \textit{Phys. stat. sol. (b)} \textbf{%
205}, 611 (1998).

\bibitem{Hg} Yu.N. Ovchinnikov and V..Z.Kresin, Phys.Rev.B \textbf{58} 12416
(1998), or J. Demsar, \textit{et al.}, Phys.Rev.B \textbf{63}, 054519 (2001).

\bibitem{chekalin} S.V.Chekalin, \textit{et al.}, Phys.Rev. Lett. \textbf{67}
3860 (1991).

\bibitem{Kochelaev2} B.I. Kochelaev, J. Sichelschmidt, B. Elschner, W.
Lemor, A. Loidl, \textit{Phys. Rev. Lett}. \textbf{79}, 4274 (1997); B.I.
Kochelaev, G.B. Tetelbaum, Superconductivity in Complex Systems, Ed.
K.A.Muller and A.Bussmann-Holder, Structure and Bonding 114, Springer
(2005), p230.,  A. Shengelaya, H. Keller, K.A. Muller, B.I. Kochelaev,
K. Conder, \textit{Phys.Rev. }\textbf{B} \textbf{63}, 144513 (2001).

\bibitem {OpticalProp}Uchida S, Ido T, Takagi H, Arima T, Tokura Y \& Tajima
S, \textit{Phys. Rev. B }
\textbf{43}, 7942(1991).

\bibitem {MattisBardeen}D.C. Mattis and J. Bardeen, \textit{Phys. Rev.
}\textbf{111}, 412 (1958).

\bibitem {Bozovic}I. Bozovic and J.N. Eckstein, \textit{Appl. Surf. Sci.}
\textbf{113/114}, 189 (1997).

\bibitem {Krasnov}V.M. Krasnov \textit{et al.}, \textit{Phys. Rev. Lett.}
\textbf{84}, 5860 (2000).


\end{thebibliography}
\end{document}